\documentclass{IOS-Book-Article}
\usepackage{times}
\usepackage{courier}
\normalfont
\usepackage[T1]{fontenc}
\usepackage[english]{babel}   
\usepackage{color}
\usepackage{graphicx}
\usepackage[utf8]{inputenc}
\usepackage[hyphens]{url}
\usepackage{amsmath}
\usepackage{amssymb}

\usepackage[bookmarks, bookmarksopen, bookmarksnumbered]{hyperref}
\usepackage[all]{hypcap}
\usepackage{listings}
\begin{document}
\begin{frontmatter}

\title{A Massive Data Parallel Computational Framework for Petascale/Exascale Hybrid Computer Systems}

\runningauthor{M. Blazewicz et al.}
\author[poznan]{\fnms{Marek} \snm{Blazewicz}}
\author[cct,cslsu]{\fnms{Steven R.} \snm{Brandt}}
\author[cct,phlsu] {\fnms{Peter} \snm{Diener}}
\author[ecelsu]{\fnms{David M.} \snm{Koppelman}}
\author[poznan]{\fnms{Krzysztof} \snm{Kurowski}}
\author[cct]{\fnms{Frank} \snm{Löffler}}
\author[pi,cct,guelph]{\fnms{Erik} \snm{Schnetter}}
\author[cct]{\fnms{Jian} \snm{Tao}}

\address[poznan]{Applications Department, Pozna\'{n} Supercomputing and Networking Center, Poland}
\address[cct]{Center for Computation \& Technology, Louisiana State University, Baton Rouge, USA}
\address[cslsu]{Department of Computer Science, Louisiana State University, Baton Rouge, USA}
\address[phlsu]{Department of Physics \& Astronomy, Louisiana State University, Baton Rouge, USA}
\address[ecelsu]{Department of Electrical and Computer Engineering, Louisiana State University, Baton Rouge, USA}
\address[pi]{Perimeter Institute for Theoretical Physics, Waterloo, Canada}
\address[guelph]{Department of Physics, University of Guelph, Guelph, Canada}

\end{frontmatter}

\thispagestyle{empty}
\pagestyle{empty}

\section*{Introduction}
Heterogeneous systems are becoming more common on High Performance Computing (HPC) systems.
Even using tools like 
CUDA~\cite{cuda40} and OpenCL~\cite{opencl11}
it is a non-trivial task to obtain optimal performance on
the GPU. Approaches to simplifying this task include 
Merge~\cite{Linderman:2008:MPM:1353536.1346318} (a library based framework
for heterogeneous multi-core systems), Zippy~\cite{CGF:CGF1131} (a framework
for parallel execution of codes on multiple GPU's), 
BSGP~\cite{Hou:2008:BBG:1360612.1360618} (a new programming language for
general purpose computation on the GPU) and 
CUDA-lite~\cite{springerlink:10.1007/978-3-540-89740-8_1} (an enhancement to
CUDA that transforms code based on annotations).
In addition, efforts are underway to improve compiler tools
for automatic parallelization and optimization of affine loop
nests for GPU's~\cite{Baskaran:2008:CFO:1375527.1375562} and for automatic
translation of OpenMP parallelized codes to 
CUDA~\cite{Lee:2009:OGC:1594835.1504194}.

In this paper we present an alternative approach: a
new computational framework
for the development of massively data parallel
scientific codes
applications suitable for use on such petascale/exascale hybrid systems built 
upon the highly
scalable Cactus framework~\cite{CS_Goodale02a, CS_Cactusweb}
As the first non-trivial
demonstration of its usefulness, we successfully developed a new 3D CFD code
that achieves improved performance.

\section{Cactus Computational Framework}
The Cactus framework~\cite{CS_Goodale02a, CS_Cactusweb} was designed
and developed to enhance programming
productivity in large-scale science collaborations.  The design of
Cactus allows scientists and engineers to develop independent components for
Cactus without worrying about portability issues on computing systems. The
common infrastructure provided by Cactus also enables developing
scientific codes that work across different disciplines. This approach emphasizes code
re-usability, and greatly simplifies constructing sound interfaces and
well-tested and well-supported software. As the name \emph{Cactus} suggests,
the Cactus framework consists of a central core called \emph{flesh}, which
provides infrastructure and interfaces for modular components called
\emph{thorns}.

Building upon the flesh, thorns can provide implementations for
computational concepts
such as parallelization,
mesh refinement, I/O, check-pointing, web servers, and so forth. The Cactus
Computational Toolkit (CCTK) is a collection of thorns which provide basic
computational capabilities. Application thorns make use of the CCTK via
the Cactus flesh API\@.
Cactus is well suited for domain discretizations via regular,
block-structured grids as are common e.g.\ for higher order finite
differences.
The \emph{Carpet} AMR library~\cite{CS_Schnetter-etal-03b,
CS_carpet_web} implements the recursive block-structured AMR
algorithm by Berger and Oliger~\cite{CS_Berger84}, and provides
support for multi-block (or multi-patch) domain discretizations.
A set of explicit time integration schemes such as
Runge-Kutta methods are provided by a Method of Lines time integrator.
Overall, the Cactus framework hides the detailed implementations of
Carpet and other utility thorns from application developers.

\subsection{MPI-Based Data Parallelism in Cactus}
The Cactus framework adopts the idea of data parallelism in its design
and implementation.
In Cactus, the computational grid is decomposed into multiple
components that are distributed between processes, and the same
set of operations are applied to each.
The communication component of Cactus uses the Message Passing Interface (MPI)
to exchange data between processes.
In Cactus, it is the task of a special \emph{driver} component to set up
storage for variables, partition the grid between MPI processes, and
manage inter-process 
communication. Unlike physical boundaries where the boundary data can be set or
calculated
from boundary conditions, data at inter-process boundaries need to be
copied from
other processes where the neighboring grid components are located.
This is implemented via a \emph{ghost region}
at the inter-process boundaries that is
automatically set up by the driver.
The necessary size of a ghost region depends on the numerical
algorithms used and can be selected as parameters at run-time.

\subsection{Parallelization on CPU-GPU Hybrid Systems}
The data parallelism in Cactus matches well with the features of
CUDA~\cite{cudasite} and OpenCL~\cite{openclsite} in supporting
programming on hybrid computer systems. On the computational framework
level, there is not much difference between CUDA and OpenCL when 
targeting NVIDIA Fermi-class GPU's. 
In this work we only focus on the parallelization on the CUDA architecture,
and will present a computational framework based on OpenCL in a later
publication.

\begin{figure}[htp]
  \centering
  \includegraphics[width=0.95\textwidth]{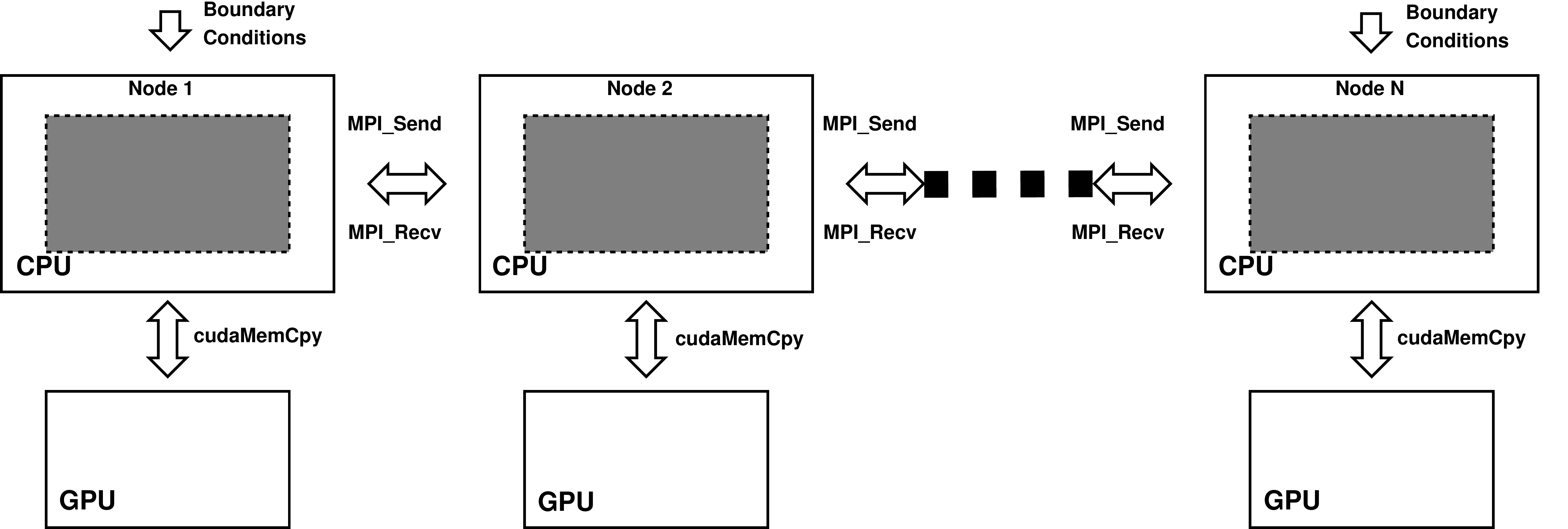}
  \caption{The Cactus computational framework manages the domain
    decomposition and 
    communication among the split domains via MPI\@. Computations are
    performed primarily on GPU's.
    The data transfers between CPU's as well as to and from the GPU's
    are concurrent with the computation.}
  \label{fig:mpicuda}
\end{figure}

Based upon the CUDA architecture, we build an MPI-CUDA based
computational framework
in Cactus~\ref{fig:mpicuda}. 
It enables a simple, semi-automatic, yet efficient implementation
and execution of CUDA-enabled applications. Auto-tuning enables
efficient data distribution
between nodes, effectively hiding additional
cost introduced by GPU-host and host-host interconnections.
The computational overhead in such a generic framework is greatly
reduced by overlapping data transfers
and computation with the asynchronous data transfers and concurrent
copy and execution supported in CUDA\@.
With the help from such a computational framework, application
developers can then spend more time
optimizing the numerical kernel itself, implementing more
efficient algorithms in these kernels, and (most importantly)
advancing the science content in their code.

This system has been tested and benchmarked on a 3D CFD implementation
(see section~\ref{CFD})
based on a finite difference discretization of Navier-Stokes equations.

\section{GPGPU Programming in Cactus}
%
%
Achieving efficient execution on a GPU often requires careful analysis
of the application followed by extensive testing and tuning. For many
important problems, such as linear algebra routines, this work has been
done and packaged into libraries for convenient use by others \cite{volkov08}.

Iterative grid techniques are widely used, and seem like a good fit
for the high floating point density of GPU's. But because each investigator
may run a different grid kernel a simple library routine would not
achieve wide use. GPU implementations of iterative grid algorithms
must deal with the problem of ghost zone exchange made more tedious by
GPU memory access constraints, among other factors.
On conventional cluster systems iterative grid application programmers
do not need to consider such issues when using a framework like
Cactus. Cactus manages data communication between a cluster's nodes,
including ghost zone exchange, so that application code need only
operate on that data. The problems related to ghost zone interchange
between CUDA blocks is similar in many ways to ghost zone interchange
between processors in a cluster CPU configuration.

In this work the Cactus framework has been extended to
cover GPU execution via an architecture neutral programming abstraction
to highly optimize finite difference operations in a multithreaded
computing environment (see list~\ref{list:kerneldef}).

\begin{figure}[htp]
  \centering
  \includegraphics[width=0.65\textwidth]{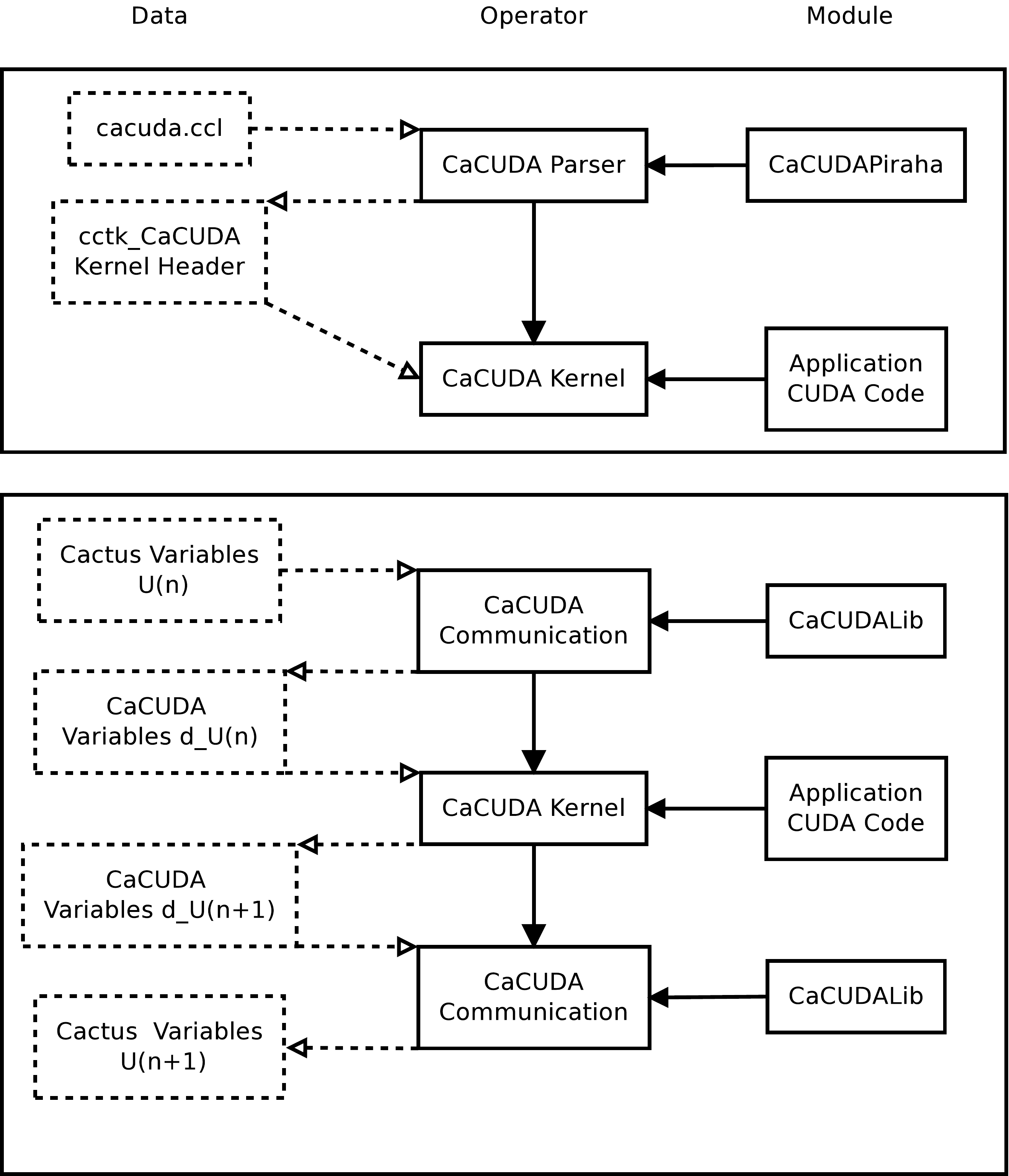}
  \caption{The workflow chart of a CaCUDA-based application. The upper box 
  shows the generation of the CaCUDA kernel headers at the code compilation stage. 
  The lower box shows how the variables are evolved to the next time step. }
  \label{fig:cacudaworkflow}
\end{figure}
\section{CaCUDA Kernel Abstraction}
The task of simplifying the generation of CUDA code for a finite differencing
code is not a straightforward one. Shared arrays with appropriate stencil sizes
have to be carefully managed, and data needed by the stencil has to be streamed
in while calculations proceed. It is possible to abstract away much of the
difficult work into boiler plate code, but doing so requires some extra
machinery. We design and implement a programming abstraction in the Cactus
framework to enable automatic generation from a set of highly optimized
templates to simplify code construction. The workflow chart of a typical
CaCUDA-based application can be found in figure~\ref{fig:cacudaworkflow}.

There are three major components in our CaCUDA Kernel abstraction.
\begin{enumerate}
\item \emph{CaCUDA Kernel Descriptor} is used to declare the variables that 
will be needed in the GPGPU computation, and identify a few relevant properties.
\item \emph{CaCUDA Templates} are a set of templates which are highly optimized 
for particular types of computational tasks and optimization strategies.
\item \emph{CaCUDA Code Generator} is used to parse the descriptors and automatically
generate CUDA-based macros. The code generator is based on Piraha\cite{Brandt10a}, 
which implements a type of parsing expression grammar\cite{Ford2004}. Due to the
page limit, we do not list the templates and the sample code generated by CaCUDA.
More about the CaCUDA project can be found at the CaCUDA project site~\cite{cacudacode}.
\end{enumerate}
\begin{lstlisting}[caption=A sample kernel definition in Cactus, label=list:kerneldef]
CCTK_CUDA_KERNEL UPDATE_VELOCITY
   TYPE=3DBLOCK
   STENCIL="1,1,1,1,1,1"
   TILE="16,16,16"
{
  CCTK_CUDA_KERNEL_VARIABLE CACHED=YES INTENT=SEPARATEINOUT 
  {
    vx, vy, vz
  } "VELOCITY"
  CCTK_CUDA_KERNEL_VARIABLE CACHED=YES INTENT=IN
  {
    p
  } "PRESSURE"
  CCTK_CUDA_KERNEL_PARAMETER
  {
    density
  } "DENSITY"
}
\end{lstlisting}
The above kernel abstraction can be integrated in a straightforward manner as a 
thorn (module), \emph{CaCUDA} in Cactus without touching the flesh (core infrastructure).
The kernel descriptor in this abstraction is similar in both format and functionality
to those Cactus Configuration Language (CCL) files, which are used to declare 
global data structures, runtime parameters, and the way various C or Fortran 
subroutines interact through the schedule tree. The abstractions already used by 
Cactus are: param.ccl, configuration.ccl, schedule.ccl, and interface.ccl. 
To this set we add an additional declarative file called cacuda.ccl.
The Cactus Framework already has a mechanism, implemented through the
configuration.ccl file, by which discovery of additional preprocessing
code can be enabled prior to the compilation of C/Fortran code.

\section{CFD Implementation}
\label{CFD}
Computational Fluid Dynamics (CFD) is one of the branches of fluid mechanics which 
uses numerical methods and algorithms to solve and analyze fluid flows. It is successfully 
used in various fields of science and engineering such as weather forecasting, 
aerodynamic optimization of body shapes (e.g. planes, cars, ships), gas reservoir uncertainty analysis. 
Unfortunately accurate CFD simulations need great computational power.
It is very important to adapt existing algorithms to new hybrid architectures and execute them
in a massively parallel manner. 

\subsection{Background and Governing Equations}
The CFD numerical method is governed by Navier-Stokes incompressible equations which are 
derived from Newton's second law (conservation of momentum) and conservation of mass 
(incompressibility). The Navier-Stokes equations are

\begin{align}\label{navier_eqs}
\frac{\partial \mathbf{u}}{\partial t} + (\mathbf{u} \cdot \mathbf{\nabla}) \mathbf{u} &= -\nabla \phi + \nu  \mathbf{\nabla}^2 \mathbf{u} + \mathbf{f}
\\
\nabla \cdot \mathbf{u} &= 0
\end{align}
where $\mathbf{u}$ is the velocity field, $\nu$ is the the kinematic viscosity, $\mathbf{f}$ is the body force,
$\phi$ is the modified pressure (pressure over density). 
The presented equations need to be further discretized in order to perform proper simulation.
In this process we have followed~\cite{VOF_Hirt79} and~\cite{NASA_VOF2D_Torrey}.
The equations are discretized using a finite-difference method. The computational 
domain is distributed onto a regular rectangular and staggered grid. 
The computations are performed in the stencil pattern. This implies that calculations 
are performed in close neighborhood of each grid's cell. 
\subsection{Code Validation and Verification}
\begin{figure}
\begin{tabular}{cc}
\includegraphics[width=0.5\textwidth]{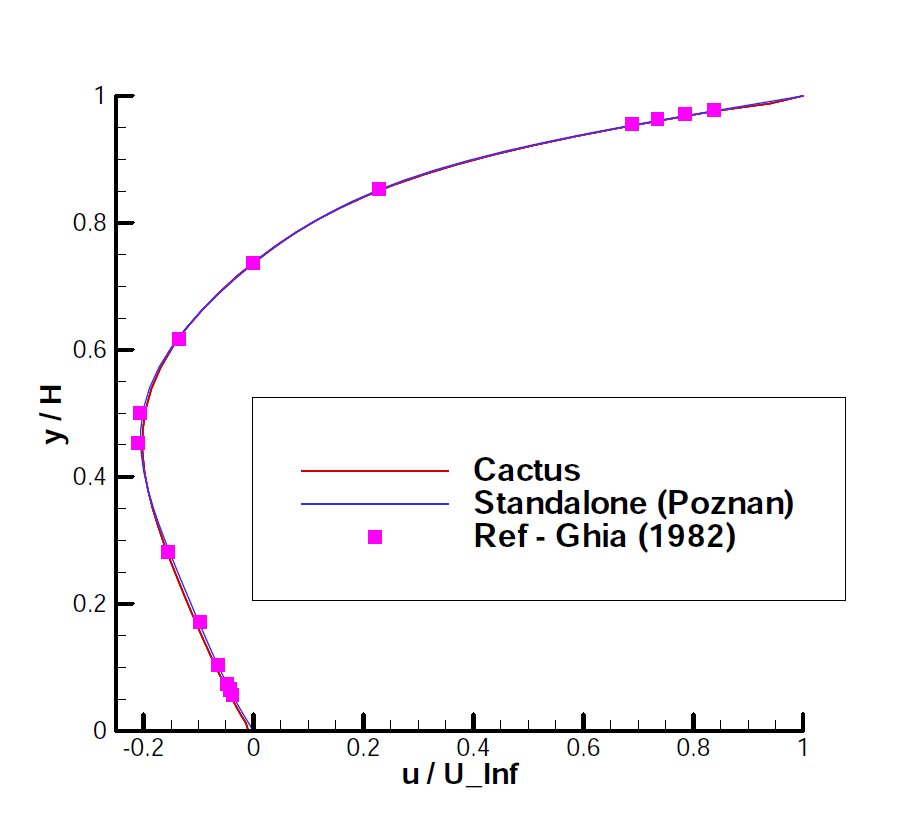}
&\includegraphics[width=0.5\textwidth]{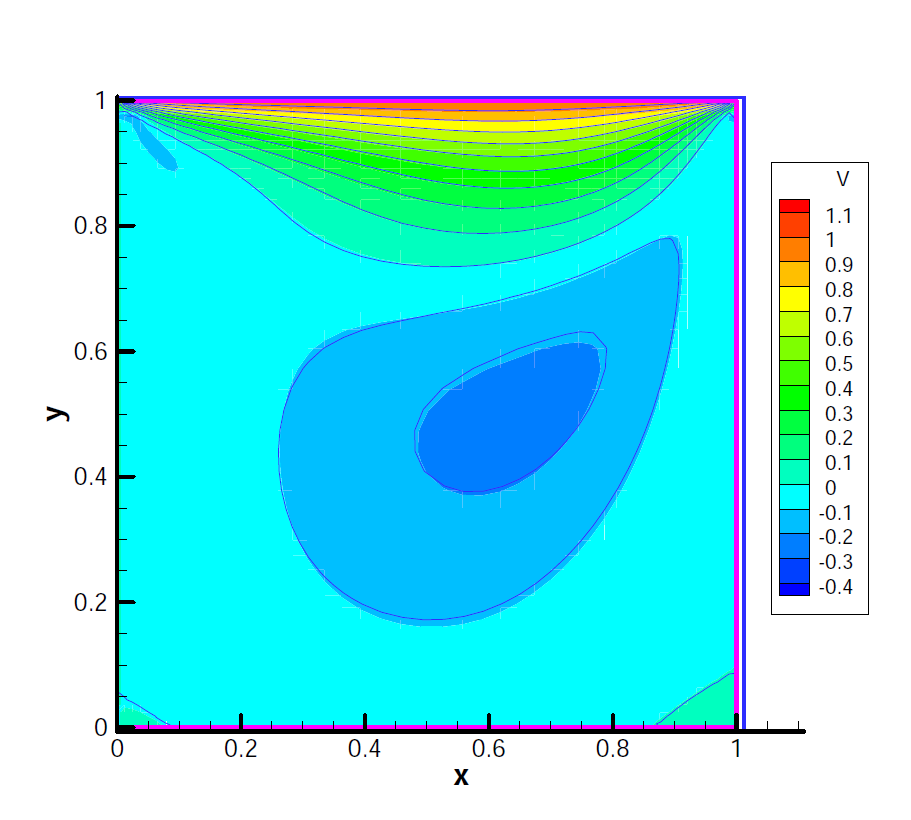}\\
\end{tabular}
\caption{The figure to the left shows the quantitative comparison of midsection centerline velocity 
with those by Ghia etc.~\cite{GhiaLDC}. The one to the right shows the contours of the X component
of the velocity field along Y axis.}
\label{fig:comparison}
\end{figure}

A homogeneous distribution of computations for the lid-driven cavity problem with a Reynolds 
number of 100 was used to benchmark the 
overall performance of the framework and verify the numerical implementation. We show 
the quantitative comparison of midsection centerline velocity with those by Ghia etc.~\cite{GhiaLDC} in
figure~\ref{fig:comparison}.

While these results come from a terascale machine, there is no logical barrier to continued scaling,
and we plan to continue scaling studies as resources become available.

\subsection{Performance and Scalability}
\begin{figure}[htp]
  \centering
  \includegraphics[width=0.9\textwidth]{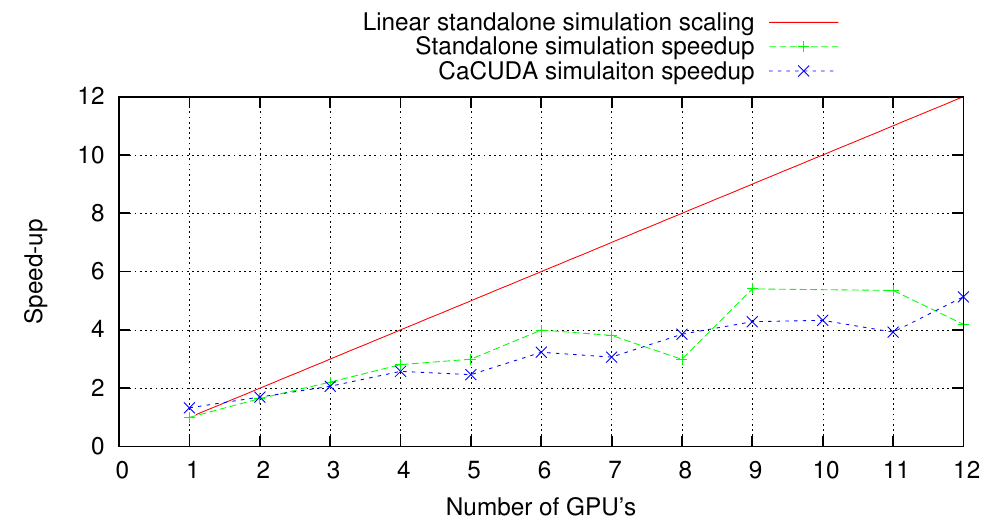}
  \caption{This plot compares the speed-up of the CFD code built with CaCUDA to 
  the standalone, handwritten implementation.
  Speed-ups are computed relative to the performance of the standalone code
  on a single node using a single GPU.}
  \label{fig:cacudacfdbench}
\end{figure}

We carried out performance and scaling tests on a 6 node GPU cluster at Cyfronet.
Each node had 2 Tesla M2050 GPU's,
two Intel Xeon X5670 processors running at 2.93GHz,
and an infiniband interconnect.
The CFD code was measured for both the standalone code and the CaCUDA
framework-based code. The performance results of one node were 43.5 and 58 GFlop/s for the 
standalone simulation and the simulation implemented within CaCUDA respectively. 
The scalability results of the 3D CFD code that makes use of the CaCUDA framework as
well as the standalone version are shown in figure~\ref{fig:cacudacfdbench}.

\section{Conclusions}
In this paper an implementation of a new generic capability for computing on hybrid CPU/GPU architectures
in the Cactus computational framework has been presented. The capability to handle the data exchange
between GPU and CPU address space and deploying the computations in the hybrid environment was
implemented as a new thorn ``CaCUDA''. Moreover the application remarkably facilitates the implementation 
process by generating the templates of all declared kernel functions.  Due to the flexibility and 
extensibility of the Cactus framework no changes to the Cactus flesh were necessary, guaranteeing 
that existing features and user implemented thorns are not affected by this addition. 

As a test case application of these new framework's features an incompressible CFD code has been 
implemented to test the overall performance and scalability. The results proving its usability 
have been presented. 

Our current effort is focused on minimizing the costs of the data exchange between GPU and CPU and 
optimizing the boundary exchange. Further integration in this area may improve performance
and scalability. 

\section*{Acknowledgments}
This work is supported by the Cybertools (http://cybertools .loni.org) project 
(NSF award 701491), the NG-CHC project (NSF award 1010640) through the Louisiana Board of Regents
, and the NSF award PIF-0904015 (CIGR).
This work used the computer resources provided by LSU/LONI.
This research was supported in part by PL-Grid Infrastructure.
This work is also supported by the UCoMS project  
under award number MNiSW(Polish Ministry of Science and Higher Education) Nr 469 1 N - USA/2009 
in close collaboration with U.S. research institutions involved
in the U.S. Department of Energy (DOE) funded grant under award number DE-FG02-04ER46136 and
the Board of Regents, State of Louisiana, under contract no. DOE/LEQSF(2004-07).
The authors want to thank our colleagues at both CCT and PSNC for great ideas
and discussions. The authors want to thank Soon-Heum Ko from the National Supercomputing Centre 
at Link\"{o}ping in Sweden for helping validating the CFD code.
 
\bibliographystyle{iopart-num}
\bibliography{CCT_CS,marqs,GPU,dmk}

\end{document}